\documentclass[12pt]{article}

\usepackage{etex}

\usepackage{amssymb,amsthm,amscd,amsbsy,array}
\usepackage{graphicx}
\usepackage{bm}
\usepackage{amsmath,dsfont}

\usepackage{graphics,graphicx,xcolor}
\usepackage[all]{xy}

\let\ssection=\section
\renewcommand{\section}{\setcounter{equation}{0}\ssection}

\newcommand{\Ad}{\mathrm{Ad}}

\newcommand{\Aff}{\mathrm{Aff}}

\newcommand{\alt}{\mathfrak{alt}}

\newcommand{\bB}{{\bf B}}

\newcommand{\bbeta}{\boldsymbol{\beta}}

\newcommand{\gammaU}{{}^{U}\!\gamma}

\newcommand{\cgal}{\mathfrak{cgal}}
\newcommand{\cmil}{\mathfrak{cmil}}
\newcommand{\sla}{\mathfrak{sl}}

\renewcommand{\d}{\mathrm{d}}

\newcommand{\rg}{g}

\newcommand{\Gal}{\mathrm{Gal}}
\newcommand{\fg}{\mathfrak{g}}

\newcommand{\CGal}{{\mathrm{CGal}}}

\newcommand{\GL}{{\mathrm{GL}}}
\newcommand{\PGL}{\mathrm{PGL}}

\newcommand{\gal}{\mathfrak{gal}}

\newcommand{\rh}{h}

\newcommand{\fh}{\mathfrak{h}}

\newcommand{\Id}{\mathrm{Id}}

\newcommand{\tLambda}{\widetilde{\Lambda}}

\newcommand{\New}{\mathrm{New}}

\newcommand{\rO}{{\mathrm{O}}}

\newcommand{\cO}{{\mathcal{O}}}

\newcommand{\bOmega}{{\boldsymbol{\Omega}}}

\newcommand{\Proj}{\mathrm{Proj}}
\newcommand{\proj}{\mathfrak{proj}}
\newcommand{\PSL}{\mathrm{PSL}}

\newcommand{\bx}{{\bm{x}}}

\newcommand{\bbR}{\mathbb{R}}

\newcommand{\Ric}{\mathrm{Ric}}
\newcommand{\RP}{\bbR\mathrm{P}}

\newcommand{\Sch}{\mathrm{Sch}}
\newcommand{\sch}{\mathfrak{sch}}

\newcommand{\SL}{\mathrm{SL}}
\newcommand{\Sl}{\mathfrak{sl}}
\newcommand{\SO}{\mathrm{SO}}
\newcommand{\Sp}{\mathrm{Sp}}

\newcommand{\so}{\mathfrak{so}}

\newcommand{\sv}{\mathfrak{sv}}

\newcommand{\Ver}{{\mathrm{Ver}}}
\newcommand{\ver}{\mathfrak{ver}}

\newcommand{\bbZ}{\mathbb{Z}}

\def\parag{\hfil\break} 
\def\kikezd{\parag\underbar}
\def\Sch{{{\rm Sch}}}

\def\Gal{{{\rm Gal}}}

\def\beq{\begin{equation}}
\def\eeq{\end{equation}}
\def\beqa{\begin{eqnarray}}
\def\eeqa{\end{eqnarray}}

\def\barray{\left(\begin{array}}
\def\earray{\end{array}\right)}

\def\SL{{\rm SL}}

\def\vx{\mathbf{x}}

\def\v0{\mathbf{0}}

\newcommand{\half }{\frac{1}{2}}

\begin{document}

\title{{\sc Conformal Galilei groups,
\\[6pt]
Veronese curves,
and
\\[6pt]
Newton-Hooke spacetimes}
\\[6pt]
}

\author{
C. DUVAL\footnote{mailto: duval-at-cpt.univ-mrs.fr}\\
Centre de Physique Th\'eorique, 
Luminy Case 907\\ 
13288 Marseille Cedex 9 (France)\footnote{ 
UMR 6207 du CNRS associ\'ee aux 
Universit\'es d'Aix-Marseille I and  II and  Universit\'e du Sud Toulon-Var; Laboratoire 
affili\'e \`a la FRUMAM-FR2291.}
 \\[12pt]
 P.~A.~HORV\'ATHY\footnote{
 On leave from the
\textit{Laboratoire de Math\'ematiques et de Physique
Th\'eorique},
 Universit\'e de Tours 
(France). mailto: horvathy-at-lmpt.univ-tours.fr}\\
 Institute of Modern Physics, Chinese Academy of Sciences
\\
Lanzhou (China)
}

\date{July 3, 2011}

\maketitle
\begin{abstract}
Finite-dimensional nonrelativistic conformal Lie algebras spanned by polynomial vector fields of Galilei spacetime arise if the dynamical exponent is $z=2/N$ with $N=1,2,\dots$. Their underlying group structure and matrix representation are constructed (up to a covering) by means of the Veronese map of degree~$N$. Suitable quotients of the conformal Galilei groups provide us with Newton-Hooke nonrelativistic spacetimes with a quantized reduced negative cosmological constant $\lambda=-N$.
\end{abstract}

\thispagestyle{empty}
\baselineskip=16pt

\bigskip
\noindent
Preprint: CPT-P006-2011

\bigskip


\newpage

\tableofcontents

\section{Introduction}

Newton-Hooke spacetimes  provide us with
solutions of the nonrelativistic gravitational field equations with nonvanishing cosmological constant, and may play a role in cosmology
 \cite{BaLL,Aldo,GiPa,Pekin,Luk07,Arratia}.  They
 can be viewed as deformations of their Galilean counterparts to which they reduce when the cosmological constant is turned off, and can indeed be obtained as  nonrelativistic limits of the de Sitter or anti-de Sitter solutions of Einstein's equations.

Another way of constructing these nonrelativistic spacetimes is to first contract the (anti-)de Sitter group to yield the ``Newton-Hooke'' group(s), and then factor out the homogeneous part of the latter \cite{GiPa}.

On the other hand, various conformal extensions of the Galilei Lie algebra have attracted much recent attention
\cite{Hen02,SZ03,LSZGalconf,Fedoruk,CGS,DHNC}, and one may wonder about their group structure and associated homogeneous spacetimes. 

This paper is devoted to studying this question.

\goodbreak

The most common, and historically first, of such extensions,
referred to as the \emph{Schr\"odinger group} \cite{Schr,DBKP-DGH,Hen94}, has been first discovered in classical mechanics~\cite{Jacobi}, and then  for the heat equation~\cite{Lie}, before being forgotten for almost a century
and then rediscovered as the maximal group of symmetries of the free Schr\"odinger equation \cite{HaPl}. 

 The Schr\"odinger group admits, in addition to those of the Galilei group, two  more generators given by their spacetime actions $(\bx,t)\mapsto(\bx^*,t^*)$, namely 
\emph{dilations}
\begin{equation}
\bx^*=a\,\bx,
\qquad
t^*=a^2t,
\label{SchDil}
\end{equation}
with $a\in\bbR^*$, and \emph{expansions} (also called \emph{inversions})
\begin{equation}
\bx^*=\Omega(t)\,\bx,
\qquad
t^*=\Omega(t)\,{t},
\label{SchExpan}
\end{equation}
where
\begin{equation}
\Omega(t)=\frac{1}{ct+1},
\label{Omega(t)}
\end{equation}
with $c\in\bbR$.

These transformations generate, along with Galilean time-translations: $\bx^*=\bx,t^*=t+b$, with $b\in\bbR$, the unimodular group~$\SL(2,\bbR)$. 
Note that the \emph{dynamical exponent} is~$z=2$; see (\ref{SchDil}).
Schr\"odinger symmetry typically arises for massive systems,
as it combines with the one-parameter central extension of the Galilei group.

The \emph{Conformal Galilei} (CG) symmetry algebra\footnote{Henkel \cite{Hen02} refers to it as 
to ``$\mathrm{Alt}_1$''.} \cite{LSZGalconf,CGS} was 
first found, and then discarded, by Barut in his attempt to derive the by then newly (re)discovered Schr\"o\-dinger symmetry  by contraction from the relativistic conformal Lie algebra~\cite{Barut}. At the group level, this new symmetry also features an
$\SL(2,\bbR)$ subgroup generated by time-translations augmented with 
modified dilations
\begin{equation}
\bx^*=a\,\bx,
\qquad
t^*=a\,t,
\label{CGDil}
\end{equation}
 and expansions
\begin{equation}
\bx^*=\Omega^2(t)\,\bx,
\qquad
t^*=\Omega(t)\,t,
\label{CGExpan}
\end{equation}
with the same parameters and factor $\Omega(t)$ as above.

This second type of 
non\-relativistic conformal symmetry 
has dynamical exponent is $z=1$, and also contains \emph{accelerations} 
\beq
\bx^*=\bx+\bB_2 t^2,
\qquad
t^*=t,
\label{accel}
\eeq
where $\bB_2\in\bbR^d$ (our notation will be justified below, see (\ref{CGalN}) and (\ref{x*})).
Moreover, this second type of conformal extension only allows for a \emph{vanishing mass} \cite{LSZGalconf}.  It is rather
difficult therefore to find physical systems 
with this kind of symmetry~\cite{CH}. 

\goodbreak

Both types of nonrelativistic symmetries have been
related to the geometric ``Newton-Cartan'' structure of nonrelativistic spacetime
\cite{HaPl,Duv0,BDP,NOR-M,DHNC}.

\goodbreak

Now, as recognized by Negro et al. \cite{NOR-M},  and by Henkel \cite{Hen97,Hen02}, both infini\-tesimal Schr\"odinger and CG symmetry belong to a much larger, generally infinite dimensional, class of Lie algebras with \emph{arbitrary},
possibly even \emph{fractional, dynamical exponent}~$z$;
 their ``conformal nonrelativistic algebra'' \cite{NOR-M} is, however
finite dimensional for the particular values \beq
z=\frac{2}{N}\,,
\qquad
N=1,2,\dots
\label{goodz}
\eeq
The terminology is justified by that, for all $z$ as in (\ref{goodz}), 
the algebra has an $\sla(2,\bbR)$ Lie subalgebra, highlighted by the dilation generator
\beq
X=\frac{1}{z}\bx\cdot\frac{\partial}{\partial\bx}+ t\,\frac{\partial}{\partial t}.
\label{zdil}
\eeq
Taking into account rotations, boosts, and translations, yields, for  $z=2$, the Schr\"odinger algebra;  the CG algebra is obtained, for $z=1$,
after incorporating also accelerations. 
 
For both $N=1$ (Schr\"odinger) and  $N=2$ (CG), the infinitesimal action integrates to a \emph{Lie group} action, but  for general $z$, the results known so far only concern \emph{Lie algebras}.  Our first new result is the derivation of the \emph{global group structure} for all~$N$ as in (\ref{goodz}).

A crucial observation for our purposes is the following: owing to the factor (\ref{Omega(t)}) in (\ref{SchExpan}) and (\ref{CGExpan}), \emph{neither} Schr\"odinger, nor Conformal Galilei transformations are \emph{globally well-defined} over ordinary Galilean spacetime.
As explained in Sections \ref{SchSection} and \ref{ConfGalGr}, Galilei spacetime should be 
replaced by a ``better one''. Our investigations in Section \ref{cosmo} show indeed that the proper arena where our conformal Galilei symmetry groups act is in fact provided by Newton-Hooke spacetimes with quantized negative cosmological constant.

\section{Nonrelativistic spacetimes
}\label{GalSchSection}

The standard \textit{Galilei spacetime} is the affine space modeled on $\bbR^{d+1}$, endowed with its cano\-nical flat affine connection $\Gamma$, and a \textit{Galilei structure} $(\gamma,\theta)$ defined by a pair of (covariantly) constant tensor fields: namely by a spatial ``metric'' and a ``clock'', expressed in an affine coordinate system $(x^1,\ldots,x^d,x^{d+1})$ as
\begin{equation}
\gamma=\sum_{i=1}^d{
\frac{\partial}{\partial x^i}\otimes\frac{\partial}{\partial x^i}
},
\qquad
\theta=\d t
\label{gamma}
\end{equation}
respectively,  where $t=x^{d+1}$ is an affine coordinate of the time axis, $T\cong\bbR$ \cite{Car,Tra,Kun,Kun1}. Notice that $\theta$ spans~$\ker(\gamma)$.

\subsection{Newton-Cartan structures}\label{NCSection}

Generalized Galilei structures consist therefore of \emph{triples} $(M,\gamma,\theta)$ where $M$ is a smooth $(d+1)$-dimensional space\-time manifold, $\gamma$ a twice-symmetric contravariant tensor field of~$M$ whose kernel is spanned by a nowhere vanishing closed $1$-form~$\theta$. 
Due to the lack of a canonical affine connection on a Galilei structure, one is compel\-led to introduce then \textit{Newton-Cartan} (NC) structures as \emph{quadruples} $(M,\gamma,\theta,\Gamma)$ where $(M,\gamma,\theta)$ is a Galilei structure, and $\Gamma$ a sym\-metric affine connection compatible with $(\gamma,\theta)$ whose curvature tensor, $R$, satisfies non-trivial extra symmetries which read, locally,
\begin{equation}
\gamma^{\mu\beta}R_{\alpha\mu\rho}^\sigma
=
\gamma^{\mu\sigma}R_{\rho\mu\alpha}^\beta
\label{SymCurv}
\end{equation}
for all $\alpha,\beta,\rho,\sigma=1,\ldots,d+1$.

Upon introducing field equations
relating the Ricci tensor to the mass-density, $\varrho$, of the sources and the cosmological constant, $\Lambda$, viz.,
\begin{equation}
\Ric=(4\pi{}G\varrho-\Lambda)\theta\otimes\theta,
\label{NCfieldEqs}
\end{equation}
the connection $\Gamma$ is interpreted as the gravitational field in a purely geometric generaliza\-tion of Newtonian gravitation theory \cite{Car,Tra,Kun}. See \cite{DBKP-DGH} for  a formula\-tion of Newton-Cartan theory in a 
Kaluza-Klein type (``Bargmann'') framework.
\goodbreak

\subsection{The Galilei group and its Lie algebra}\label{GalSection}

The 
\textit{Galilei group}, $\Gal(d)$, consists of all diffeomorphisms $\rg$  
of space-time which preserve all three ingredients of the Galilei structure, i.e., such that 
\beq
\rg_*\gamma=\gamma,
\quad
\rg_*\theta=\theta,
\quad
\rg_*\Gamma=\Gamma.
\label{GalDef}
\eeq
This is the group of symmetries that governs nonrelativistic physics in $d$ spatial dimensions. It clearly consists of $(d+2)\times(d+2)$ matrices of the form \cite{Sou}
\begin{equation}
\rg=
\left(
\begin{array}{ccc}
A&\bB_1&\bB_0\\
0&1&b\\
0&0&1
\end{array}
\right)\in\Gal(d),
\label{Gal}
\end{equation}
where $A\in\rO(d)$, and $\bB_0,\bB_1\in\bbR^d$ stand respectively for a space translation and a boost, and $b\in\bbR$ is a time translation. 
The (affine) action of $\Gal(d)$ on spacetime~$\bbR^d\times\bbR$ reads
\begin{equation}
\rg_{\bbR^{d+1}}:
\left(
\begin{array}{c}
\bx\\
t\\
1
\end{array}
\right)
\mapsto
\left(
\begin{array}{c}
A\bx+\bB_1 t+\bB_0\\
t+b\\
1
\end{array}
\right).
\label{GalAction}
\end{equation}
Infinitesimal Galilei transformations form hence a Lie algebra, $\gal(d)$, spanned by all vector fields $X$ on space-time such that 
\beq
L_X\gamma=0,\quad
L_X\theta=0, \quad
L_X\Gamma=0
\eeq
(see~\cite{Tra,Duv0} for a generalization to (curved) NC structures); these vector field read
\begin{equation}
X=
\left(\omega^i_jx^j+\beta_1^i t+\beta_0^i\right)\frac{\partial}{\partial x^i}
+\varepsilon\frac{\partial}{\partial t},
\label{gal} 
\end{equation}
where $\omega\in\so(d)$, $\bbeta_0,\bbeta_1\in\bbR^d$, and $\varepsilon\in\bbR$. Latin indices run in the range $1,\dots,d$, and Einstein's summation convention is assumed throughout this article.

The Lie algebra~$\gal(d)$ admits the faithful $(d+2)$-dimensional anti-representation $X\mapsto{}Z$ where,
\begin{equation}
Z=
\left(
\begin{array}{ccc}
\omega&\bbeta_1&\bbeta_0\\
0&0&\varepsilon\\
0&0&0
\end{array}
\right)\in\gal(d)
\label{galalRep} 
\end{equation}
with the above notation. 

\subsection{The Schr\"odinger group and its Lie algebra}\label{SchSection}

Let us first discuss the 
\textit{Schr\"odinger group}, $\Sch(d)$, which  includes, in addition to the standard Galilei generators, those of the projective group, $\PSL(2,\bbR)$, of the time axis.
Up to a quotient that we will make more precise later on, the Schr\"odinger group will be defined as the matrix group whose typical element reads \cite{Per,Duv0}
\begin{equation}
\rg=
\left(
\begin{array}{ccc}
A&\bB_1&\bB_0\\
0&a&b\\
0&c&d
\end{array}
\right)\in\Sch(d),
\label{SchGroup}
\end{equation}
where $A\in\rO(d)$, $\bB_0,\bB_1\in\bbR^d$, and $a, b, c, d\in\bbR$ with $ad-bc=1$.
The \textit{projective} ``action'' of $\rg\in\Sch(d)$ on spacetime~$\bbR^d\times\bbR$ takes the form
\begin{equation}
\rg_{\bbR^{d+1}}:
\left(
\begin{array}{c}
\bx\\
t\\
1
\end{array}
\right)
\mapsto
\left(\begin{array}{c}
\displaystyle
\frac{A\bx+\bB_1 t+\bB_0}{ct+d}\\[10pt]
\displaystyle
\frac{at+b}{ct+d}\\[10pt]
1
\end{array}
\right)
\label{SchAction}
\end{equation}
defined on the \emph{open subset of spacetime} where $ct+d\neq0$.

\goodbreak
It is an easy matter to check that the action 
(\ref{SchAction}) is consistent with the one presented in the introduction; Schr\"odinger dilations (\ref{SchDil}) correspond
to $b=0, c=0$,
and expansions (\ref{SchExpan}) to $a=1,b=0,d=1$.

The group structure is 
$\Sch(d)=(\rO(d)\times\SL(2,\bbR))\ltimes(\bbR^d\times\bbR^d)$.  

Now,
in order to guarantee a well-behaved action of this group on spacetime, one must demand that \emph{time be compactified}, $T\cong\RP^1$.
In fact, the Schr\"odinger group does \emph{not} act on ``ordinary'' Galilei spacetime, but rather on the \textit{M\"obius manifold}
\begin{equation}
M=\big(\bbR^d\times(\bbR^2\!\setminus\!\{0\})\big)/\bbR^*
\label{M1}
\end{equation}
fibered above the projective line, $\RP^1$, as clear from (\ref{SchAction}). This point will be further developed in Section \ref{CGalSpaceTimeSection}.
See also \cite{Duv0}.

Note that 
 (\ref{M1}) can be recovered by factoring out the
homogeneous subgroup generated by rotations, expansions, dilations, and boosts,
\beq
M=\Sch(d)/H
\quad\hbox{where}\quad
H=\big(\rO(d)\times\Aff(1,\bbR)\big)\ltimes\bbR^d,
\eeq
where $\Aff(1,\bbR)$ stands for the $2$-dimensional group of lower-triangular matrices in $\SL(2,\bbR)$, generated by dilations and expansions.

Note that,
unlike conformally compactified Minkowski spacetime $(S^d\times{}S^1)/\bbZ_2$, only time, not space, is compactified here since
\beq
 M\cong(\bbR^d\times{}S^1)/\bbZ_2.
 \eeq
 
It will be shown in Section \ref{cosmo} that the
M\"obius manifold carries a nonrelativistic Newton-Cartan structure; it is, in fact a \emph{Newton-Hooke spacetime with cosmological constant} $\Lambda=-d$, minus the dimension of space; see (\ref{Lambda1}).
 
The Schr\"odinger group can, indeed, be defined in a geometric way, namely in the NC framework  \cite{Duv0,DBKP-DGH,DHNC}, as the group, $\Sch(d)$, of all (locally defined) diffeomorphisms $\rg$ 
 such that 
\begin{equation}
\rg_*(\gamma\otimes\theta)=\gamma\otimes\theta
\qquad
\&
\qquad
\rg\in\Proj(\bbR^{d+1},\Gamma),
\label{SchDef}
\end{equation}
where $\Proj(\bbR^{d+1},\Gamma)$ denotes the set of all projective transformations of spacetime, namely of all (local) diffeomorphisms which permute the geodesics of spacetime w.r.t. the connection $\Gamma$. 
Let us stress that the conditions (\ref{SchDef}) imply, in particular, that the diffeo\-morphism~$\rg$ projects on the time axis as an element of $\PGL(2,\bbR)$ which must also preserve time-orientation defined by $\theta$, namely an element of $\PSL(2,\bbR)$. The general solution of~(\ref{SchDef}) is therefore given by (\ref{SchGroup}), up to a covering; see also~(\ref{SchAction}).

\goodbreak

The {Schr\"odinger Lie algebra}, $\sch(d)$, is then the Lie algebra of those vector fields $X$ on spacetime  such that 
 \begin{equation}
L_X(\gamma\otimes\theta)=0
\qquad
\&
\qquad
X\in\proj(\bbR^{d+1},\Gamma).
\label{schDef}
\end{equation}
In local terms, we thus require 
\beq
L_X\gamma^{\alpha\beta}\theta_\rho
+
\gamma^{\alpha\beta}L_X\theta_\rho=0
\qquad
\&
\qquad
L_X\Gamma^\rho_{\alpha\beta}=\delta^\rho_\alpha\varphi_\beta+\delta^\rho_\beta\varphi_\alpha
\eeq
 for some $1$-form $\varphi$ of $\bbR^{d+1}$ depending on $X$, and for all $\alpha,\beta,\rho=1,\ldots,d+1$.

We easily find that $X\in\sch(d)$ iff
\begin{equation}
X
=
\left(\omega^i_j\,x^j+\kappa{}t{}x^i+\lambda{}x^i+\beta_1^i{}t+\beta_0^i\right)\frac{\partial}{\partial x^i}
+\left(\kappa{}t^2+2\lambda{}t+\varepsilon\right)\frac{\partial}{\partial t}\,,
\label{schd} 
\end{equation}
where $\omega\in\so(d)$, $\bbeta_0,\bbeta_1\in\bbR^d$, and $\kappa,\lambda,\varepsilon\in\bbR$. 
 The Schr\"odinger dilation (or homothety) generator is, indeed, 
(\ref{zdil}) with dynamical exponent $z=2$.

The Lie algebra $\sch(d)$ admits the faithful $(d+2)$-dimensional anti-representa\-tion 
$X\mapsto{}Z$, 
where
\begin{equation}
Z=
\left(
\begin{array}{rrr}
\omega&\bbeta_1&\bbeta_0\\[6pt]
0&\lambda&\varepsilon\\[6pt]
0&-\kappa&-\lambda
\end{array}
\right)\in\sch(d)
\label{sch2dRep} 
\end{equation}
with the same notation as above. 

Note that $\sch(d)$ is, in fact, the (centerless) Schr\"odinger Lie algebra. Physical applications also involve a central extension associated with the mass; see, e.g., \cite{Duv0,BDP,Schr,DBKP-DGH,Hen94,Hen02,DHNC}.

A remarkable property of the Schr\"odinger group, 
arising as a symmetry group of the classical space of motions of free spinning particle \cite{DHNC}, and also important
for studying supersymmetric extensions \cite{SchSUSY},
 is that it
can  be faithfully imbedded into the affine-symplectic Lie algebra
\beq
\Sch(d)\subset\Sp(d,\bbR)\ltimes\bbR^{2d}.
\eeq

\section{Conformal Newton-Cartan transformations
\& 
finite-dimensional conformal Galilei Lie algebras}\label{CNCSection}

In close relationship with the Lorentzian framework, we call \textit{conformal Galilei} transformation of a general Galilei spacetime $(M,\gamma,\theta)$ any diffeomorphism of $M$ that preserves the direction of~$\gamma$. Owing to the fundamental constraint $\gamma(\theta)=0$, it follows that conformal Galilei transformations automatically preserve the direction of the time $1$-form~$\theta$.

In terms of infinitesimal transformations, a \textit{conformal Galilei} vector field of $(M,\gamma,\theta)$ is a vector field, $X$, of $M$ that Lie-transports the direction of $\gamma$; we will thus define $X\in\cgal(M,\gamma,\theta)$ iff
\begin{equation}
L_X\gamma=f\gamma
\qquad
\hbox{hence}
\qquad
L_X\theta=g\,\theta
\label{confgal} 
\end{equation}  
for some smooth functions $f,g$ of $M$, depending on $X$. Then, $\cgal(M,\gamma,\theta)$ becomes a Lie algebra whose bracket is the Lie bracket of vector fields.

The one-form $\theta$ being parallel-transported by the NC-connection, one has neces\-sarily $d\theta=0$; this yields $dg\wedge\theta=0$, implying that $g$ is (the pull-back of) a smooth function on~$T$, i.e., that $g(t)$ depends arbitrarily on time $t=x^{d+1}$, which locally parametrizes the time axis. We thus have $dg=g'(t)\theta$.

\subsection{Conformal Galilei transformations, $\cgal_{2/z}(d)$, with dyna\-mical ex\-ponent~$z$}\label{cgalzSection}

One can, at this stage, try and seek nonrelativistic avatars of general relativistic infinitesimal conformal transformations. Given a Lorentzian (or, more generally, a pseudo-Riemannian) manifold $(M,\rg)$, the latter Lie algebra is generated by the vector fields, $X$, of~$M$ such that
\begin{equation}
L_X(\rg^{-1}\otimes\rg)=0,
\label{conf} 
\end{equation}
where $\rg^{-1}$ denotes the inverse of the metric $\rg:TM\to{}T^*M$.

It has been shown \cite{Kun1} that one can expand a Lorentz metric in terms of the small parameter $1/c^2$, where $c$ stands for the speed of light, as  
\beq
\rg=c^2\theta\otimes\theta-\gammaU+\cO(c^{-2}), 
\qquad
\rg^{-1}=-\gamma+c^{-2}U\otimes{}U+\cO(c^{-4}),
\label{limg}
\eeq
with the previous notation. Here $U$ is an ``observer'', i.e., a smooth timelike vector field of spacetime $M$, such that $\rg(U,U)=c^2$, around which the light-cone opens up in order to consistently define a procedure of nonrelativistic limit. The Galilei structure~$(\gamma,\theta)$ is recovered via $\gamma=-\lim_{c\to\infty}\rg^{-1}$, and $\theta=\lim_{c\to\infty}(c^{-2}\rg(U))$. In~(\ref{limg}) the symmetric twice-covariant tensor field $\gammaU$ will define the Riemannian metric of the spacelike slices in the limiting Galilei structure.

We can thus infer that the nonrelativistic limit of Equation (\ref{conf}) would be $L_X\lim_{c\to\infty}(c^{-2}\,\rg^{-1}\otimes\rg)=0$, viz.
\begin{equation}
L_X(\gamma\otimes\theta\otimes\theta)=0.
\label{q=2} 
\end{equation}

\goodbreak

More generally, we  consider
\begin{equation}
L_X(\gamma^{\otimes{}m}\otimes\theta^{\otimes{}n})=0,
\label{galconfMN} 
\end{equation}
for some $m=1,2,3,\ldots$, and $n=0,1,2,\ldots$, to be further imposed on the vector fields $X\in\cgal(M,\gamma,\theta)$. Then the quantity
\begin{equation}
z=\frac{2}{q}
\qquad\hbox{where}\qquad
q=\frac{n}{m}\,
\label{z} 
\end{equation}
matches the ordinary notion of dynamical exponent \cite{Hen94,Hen02,DHNC}.

 We will, hence, introduce the Galilean avatars, $\cgal_{2/z}(M,\gamma,\theta)$, of the Lie algebra~$\so(d+1,2)$ of conformal vector fields of a pseudo-Riemannian structure of signature $(d,1)$ as the Lie algebras spanned by the vector fields $X$ of $M$ satisfying~(\ref{confgal}), and (\ref{galconfMN}).
 We will call $\cgal_{2/z}(M,\gamma,\theta)$ the \textit{conformal Galilei Lie algebra with dynamical exponent} $z$ in (\ref{z}). This somewhat strange notation will be justified in the sequel.

The Lie algebra 
\begin{equation}
\sv(M,\gamma,\theta)=\cgal_1(M,\gamma,\theta)
\label{sv}
\end{equation}
is the obvious generalization to Galilei spacetimes of the \textit{Schr\"odinger-Virasoro} Lie algebra $\sv(d)=\sv(\bbR\times\bbR^d,\gamma,\theta)$ introduced in \cite{Hen94} (see also \cite{Hen02}) from a different viewpoint in the case of a flat NC-structure.  The representations of the Schr\"odinger-Virasoro group and of its Lie algebra, $\sv(d)$, as well as the deformations of the latter have been thoroughly studied and investigated in \cite{RU,RU2}. 

 Let us henceforth use the notation $\cgal_{2/z}(d)=\cgal_{2/z}(\bbR^{d+1},\gamma,\theta)$ with $\gamma$  as in (\ref{gamma}) and and~$\theta=dt$ respectively.
Then one shows \cite{DHNC}
 that $X\in\cgal_{2/z}(d)$ iff
\begin{equation}
X=
\Big(\omega^i_j(t)x^j+\frac{1}{z}\xi'(t)x^i+\beta^i(t)\Big)\frac{\partial}{\partial x^i}
+\xi(t)\frac{\partial}{\partial t}\,,
\label{cgalz} 
\end{equation}
where $\omega(t)\in\so(d)$, $\bbeta(t)$, and $\xi(t)$ depend arbitrarily on time, $t$. 

The Lie algebra $\cgal_{0}(M,\gamma,\theta)$ corresponding to the case $z=\infty$ is also interesting; it is a Lie algebra of symplectomorphisms of the models of massless and spinning  Galilean particles \cite{Duv0,DHNC}.

\subsection{The Lie algebra, $\cgal_{N}(d)$, of finite-dimensional conformal Galilei transformations}
\label{cgal2/NSection}

Now we show  that our formalism leads to a natural definition of a whole family of distinguished finite-dimensional Lie subalgebras of the conformal Galilei Lie algebra $\cgal_{2/z}(d)$ with prescribed dynamical exponent $z$,
generated by the vector fields in~(\ref{cgalz}),
where $\omega(t)\in\so(d)$, $\bbeta(t)$, and $\xi(t)$ depend smoothly on time, $t$. 

Referring to \cite{DHNC} for details,
let us restrict our attention to those vector fields $X\in\cgal^\mathrm{Pol}_{2/z}(d)$ that are \emph{polynomials} of fixed degree $N>0$ in the variables
 $x^1, \ldots, x^d$, and $t=x^{d+1}$.
We then have necessarily
\beq
\xi(t)=\kappa t^2+2\lambda t+\varepsilon,
\label{xi2}
\eeq
with $\kappa,\lambda,\varepsilon\in\bbR$, 
and we find that a closed Lie algebra of polynomial vector fields of degree $N>0$ is obtained provided
\begin{equation}
z=\frac{2}{N}.
\label{z=2/N}
\end{equation}
At last, we find that $X\in\cgal^\mathrm{Pol}_{N}(d)$ iff
\beq
X=
\Big(\omega^i_j x^j+\frac{N}{2}\xi'(t)x^i+\beta^i(t)\Big)\frac{\partial}{\partial{}x^i}
+
\xi(t)\frac{\partial}{\partial t}, 
\label{cgalN}
\eeq
with $\omega\in\so(d)$, and $\xi(t)$ quadratic as in (\ref{xi2}), together with
\beq
\bbeta(t)=\bbeta_N t^N+\cdots+\bbeta_1 t+\bbeta_0,
\eeq
where $\bbeta_0,\ldots\bbeta_N\in\bbR^d$.
The finite-dimensional Lie algebras $\cgal^\mathrm{Pol}_{N}(d)$ 
turn out to be isomorphic to the so-called $\alt_{2/N}(d)$ Lie algebras discovered by Henkel \cite{Hen94} in his study of scale invariance for strongly anisotropic critical systems (with $d=1$), viz.,
$
\cgal^\mathrm{Pol}_{N}(d)\cong\alt_{2/N}(d).
$ 

 From now on we drop the superscript ``Pol'' as no further confusion can occur. In the case $N=1$, we recognize the Schr\"odinger Lie algebra $\cgal_{1}(d)\cong\sch(d)$, see~(\ref{schd}),\footnote{Strictly speaking, for the lowest level, $N=1$, the vector fields generating $\cgal_1(d)$ are poly\-nomials of degree $2$ in the \textit{spacetime} coordinates, although $z=2$ holds true. The higher levels $N\geq2$ duly correspond to the actual degree of the vector fields generating~$\cgal_N(d)$.} while for $N=2$ we recover the ``Conformal Galilei Algebra'' (CGA) $\cgal_{2}(d)$, called  $\cmil_1(d)$ in \cite{DHNC}.

\section{Conformal Galilei Groups with dyna\-mical exponents $z=2/N$}\label{ConfGalGr}

\subsection{Veronese curves and finite-dimensional representations of $\SL(2,\bbR)$}

A \textit{Veronese curve} is an embedding $\Ver_N:\RP^1\to\RP^N$ defined, for $N\geq1$ by
\begin{equation}
\Ver_N(t_1:t_2)=(t_1^N:t_1^{N-1}t_2:\cdots:t_2^N),
\label{Ver}
\end{equation}
where $(u_1:u_2:\cdots:u_{N+1})$ stands for the direction of 
$(u_1,u_2,\ldots,u_{N+1})\in\bbR^{N+1}\!\setminus\!\{0\},$
 that is, a point in $\RP^N$. See, e.g., \cite{OT}.
 
\goodbreak

With a slight abuse of notation, we will still denote by $\Ver_N:\bbR^2\to\bbR^{N+1}$ the mapping defined by
\begin{equation}
\Ver_N(t_1,t_2)=(u_1,u_2,\ldots,u_{N+1})
\qquad
\text{where}
\qquad 
u_k=t_1^{N-k+1}t_2^{k-1}
\label{uk}
\end{equation}
for all $k=1,\ldots,N+1$.
Put $t=(t_1,t_2)\in\bbR^2$, and consider $t^*=Ct$ with 
\begin{equation}
C=
\left(\begin{array}{cc}
a&b\\c&d
\end{array}
\right)\in\SL(2,\bbR).
\label{SL2R}
\end{equation}
The image~$u^*$ of~$t^*$ under the Veronese map is clearly a $(N+1)$-tuple of homogeneous polynomials of degree~$N$ in $t$; it thus depends linearly on $u=(u_1,\ldots,u_{N+1})\in\bbR^{N+1}$, where the $u_k$ are as in (\ref{uk}). 
The general formula is as follows. If $t_1^*=at_1+bt_2$, $t_2^*=ct_1+dt_2$, with $ad-bc=1$, then 
\beq
\Ver_N(Ct)=\Ver_N(C)\Ver_N(t),
\eeq
 where $\Ver_N(C)$ a nonsingular $(N+1)\times(N+1)$ matrix with entries
\begin{eqnarray}
\nonumber
\Ver_N(C)^m_{m'}&=&
\sum_{k=\max(0,m'-m)}^{\min(N-m+1,m'-1)}
\left(
\begin{array}{c}
N-m+1\\
k
\end{array}
\right)
\left(
\begin{array}{c}
m-1\\
m'-k-1
\end{array}
\right)\times\\[6pt]
&&\times{}a^{N-m-k+1}b^k c^{m-m'+k} d^{m'-k-1}
\label{VNC}
\end{eqnarray}
for all $m,m'=1,\ldots,N+1$.
Our mapping  provides us with a group homomorphism
\begin{equation}
\Ver_N:\SL(2,\bbR)\to\SL(N+1,\bbR)
\label{HomVer}
\end{equation}
which constitutes (up to equivalence) the well-known $(N+1)$-dimensional irreducible representation of $\SL(2,\bbR)$; see \cite{Kna}.
Let us introduce
the $\Sl(2,\bbR)$ generators
\begin{equation}
\xi_{-1}
=\left(\begin{array}{cc}
0&1\\0&0
\end{array}
\right),
\qquad
\xi_0
=\left(\begin{array}{cc}
1&0\\0&-1
\end{array}
\right),
\qquad
\xi_{1}
=\left(\begin{array}{cc}
0&0\\1&0
\end{array}
\right),
\label{sl2matrix}
\end{equation}
interpreted physically as the infinitesimal generators of 
time translations 
$\xi_{-1}$,
dilations $\xi_0$, and  expansions $\xi_1$. 
Their images under the tangent map of of $\Ver_N$ at the identity read then
\begin{eqnarray}
\label{verY}
\ver_N(\xi_{-1})&=&\sum_{n=1}^{N+1}(N-n+1)\,u_{n+1}\frac{\partial}{\partial{u_n}},
\\[6pt]
\label{verZ}
\ver_N(\xi_0)&=&\sum_{n=1}^{N+1}{(N-2n+2)\,u_{n}\frac{\partial}{\partial{u_n}}},
\\[6pt]
\label{verX}
\ver_N(\xi_{1})&=&\sum_{n=1}^{N+1}{(n-1)\,u_{n-1}\frac{\partial}{\partial{u_n}}}.
\end{eqnarray}
 One checks that $\ver_N(\xi_{a})$ is, indeed, divergence-free,  and 
\beq
[\ver_N(\xi_a),\ver_N(\xi_b)]=-\ver_N([\xi_a,\xi_b])
\eeq
for $a,b=-1,0,1$, i.e., that $\ver_N:\Sl(2,\bbR)\to\Sl(N+1,\bbR)$ is
a Lie algebra anti-homomorphism.

\goodbreak

\subsection{Matrix realizations of the Conformal Galilei Groups $\mathbf{CGal}_N(d)$}

Just as in the case of the Schr\"odinger group, see (\ref{SchGroup}), we will strive integrating the conformal Galilei Lie algebras $\cgal_{N}(d)$ within the matrix group $\GL(d+N+1,\bbR)$. 
Let us, hence, introduce the \textit{Conformal Galilei Group} with \textit{dynamical exponent} $z={2}/{N}$ 
cf. (\ref{z=2/N}),
which we denote by $\CGal_{N}(d)$; it consists of those matrices of the form
\begin{equation}
\rg=
\left(
\begin{array}{cc}
A&\begin{array}{ccc}
\bB_N\!\!\!&\cdots&\!\!\!\bB_0
\end{array}
\\[7pt]
0&\Ver_N(C)
\end{array}
\right)
\label{CGalN}
\end{equation}
where $A\in\rO(d)$, $\bB_0,\bB_1,\ldots,\bB_N\in\bbR^d$, and $C\in\SL(2,\bbR)$. 

We now prove that the Lie algebra of $\CGal_{N}(d)$ is, indeed, $\cgal_{N}(d)$ introduced in Section \ref{CNCSection}. 
In fact, putting $t=t_1/t_2$ in (\ref{uk}), wherever $t_2\neq0$, we easily find that the projective action $\rg_{\bbR^{d+1}}:(\bx,t)\mapsto(\bx^*,t^*)$ of $\CGal_{N}(d)$ reads, locally, as
\begin{equation}
\barray{c}
\bx^*\\ t_*^N\\ \vdots\\ t^*\\ 1
\earray
=
\bbR^*\cdot\rg
\barray{c}
\bx\\ t^N\\ \vdots\\ t\\ 1
\earray,
\label{projAction}
\end{equation}
which, with the help of (\ref{SL2R}) and (\ref{VNC}), leaves us with
\begin{eqnarray}
\label{x*}
\bx^*&=&\frac{A\bx+\bB_N t^N+\cdots+\bB_1 t+\bB_0}{(c t+d)^N},\\[4pt]
\label{t*}
t^*&=&\frac{a t + b}{c t + d}.
\end{eqnarray}
These formul{\ae} allow for the following interpretation for the  parameters in~(\ref{CGalN}), 
\beq
\begin{array}{rl}
A: &\hbox{orthogonal transformation},
\\
\bB_0: &\hbox{translation},
\\
\bB_1: &\hbox{boost
},
\\
\bB_2: &\hbox{acceleration},
\\
\vdots &\vdots\\
\bB_N: &\hbox{higher-order 
``acceleration''},
\\
C: &\text{projective transformation of time}.
\end{array}
\label{Naccel}
\eeq

\goodbreak
Let us now write any vector in the Lie algebra of
$\CGal_{N}(d)$ as 
\begin{equation}
Z=
\left(
\begin{array}{cc}
\omega&\begin{array}{ccc}
\bbeta_N\!\!\!&\cdots&\!\!\!\bbeta_0
\end{array}
\\[7pt]
0&\ver_N(\xi)
\end{array}
\right),
\label{LieCGalN}
\end{equation}
where we have used Equations (\ref{verY})--(\ref{verX}) with
\begin{equation}
\xi=
\left(
\begin{array}{rr}
\lambda&\varepsilon\\[6pt]
-\kappa&-\lambda
\end{array}
\right)\in\Sl(2,\bbR).
\label{sl2R} 
\end{equation}
Then the infinitesimal form of the transformation laws (\ref{x*}) and (\ref{t*}) writes as $
\delta{\bx}=\delta{\bx^*}\strut\vert_{\delta\rg=Z,\rg=\Id}$, together with $\delta{}t=\delta{t^*}\strut\vert_{\delta\rg=Z,\rg=\Id}$, i.e., 
\begin{equation}
\delta{\bx} =\omega\bx+\bbeta_N{}t^N+\cdots\bbeta_1{}t+\bbeta_0+N(\kappa{}t+\lambda)\bx,
\qquad
\delta{t} =\kappa{}t^2+2\lambda{}t+\varepsilon.
\end{equation}
At last, the vector field $Z_{\bbR^{d+1}}=\delta{x}^i\,\partial/\partial{x^i}+\delta{t}\,\partial/\partial{t}$ associated with $Z$ in (\ref{LieCGalN}) is such that
\begin{equation}
Z_{\bbR^{d+1}}=X\in\cgal_{N}(d),
\end{equation}
where the vector field $X$ is as in (\ref{cgalN}), proving our claim.

Our terminology for the dynamical exponent is justified  by verifying that the dilation generator is (\ref{zdil}) with $z=2/N$.

The above definition of the conformal Galilei groups, see (\ref{CGalN}), yield their global structure 
\begin{equation}
\CGal_{N}(d)\cong(\rO(d)\times\SL(2,\bbR))\ltimes\bbR^{(N+1)d},
\label{LeviCGalN}
\end{equation}
and $\dim(\CGal_{N}(d))=Nd+\half{}d(d+1)+3$. 

$\bullet$ For $N=1$, i.e., $z=2$, we recover the Schr\"odinger group (\ref{SchGroup}), and therefore 
\beq
\CGal_{1}(d)\cong\Sch(d).
\eeq

\goodbreak

$\bullet$ For $N=2$, i.e., $z=1$, in particular, we get 
\begin{equation}
\rg=
\left(
\begin{array}{cccc}
A&\bB_2&\bB_1&\bB_0\\
0&a^2&2ab&b^2\\
0&ac&ad+bc&bd\\
0&c^2&2cd&d^2
\end{array}
\right)\in\CGal_{2}(d),
\label{CGal2}
\end{equation}
with $A\in\rO(d)$, $\bB_0,\bB_1,\bB_2\in\bbR^d$, $a,b,c,d\in\bbR$ with $ad-bc=1$. 
In addition to the usual space translations $\bB_0$, and Galilei boosts~$\bB_1$, we also have extra generators, namely \textit{accelerations} $\bB_2$ \cite{LSZGalconf,CGS}.
It is an easy matter to check that the actions~(\ref{CGDil}), (\ref{CGExpan}), and (\ref{accel}) of dilations, expansions, and accelerations, respectively, are recovered by considering their projective action given by (\ref{projAction}) with $N=2$.

\goodbreak

The Lie algebra $\cgal_{2}(d)$ of the Conformal Galilei Group, $\CGal_{2}(d)$, is plainly isomorphic to the (centerless) 
Conformal Galilei Algebra (CGA), cf.  \cite{DHNC,LSZGalconf,CGS}.
It has been shown \cite{LSZGalconf} that $\cgal_{2}(d)$ admits a nontrivial $1$-dimensional central extension in the planar case, $d=2$, only.

\section{Conformal Galilei spacetimes \& cosmo\-logical constant}\label{cosmo}

As mentioned in the Introduction, the physical spacetime can be recovered by postulating some symmetry group  --- defining its geometry --- and then by factoring
out a suitable subgroup.  The simplest example is 
to start with the neutral component of the Galilei group (\ref{Gal}), namely
\begin{equation}
\Gal_+(d)=(\SO(d)\times\bbR)\ltimes\bbR^{2d},
\end{equation}
and
factor out rotations and  boosts to yield
(ordinary) Galilei spacetime,
\beq
\bbR^d\times\bbR=\Gal_+(d)/(\SO(d)\ltimes\bbR^d).
\label{GalST}
\eeq
Similarly, one can start instead with a deformation of the Galilei group called Newton-Hooke group  
\cite{BaLL,Aldo,GiPa}
\beqa
N^+(d)=\big(\SO(d)\times\SO(2)\big)\ltimes\bbR^{2d},
\label{NH+}
\eeqa
where $\SO(d)\times\SO(2)$ 
is the direct product of spatial rotations and translations of (compactified) time
acting on the Abelian subgroup~$\bbR^{2d}$ of
boosts and space-translations.
Then, quotienting $N^{+}(d)$ by the direct product of rotations and boosts yields the \emph{Newton-Hooke spacetime} \cite{GiPa}. The latter carries a  non-flat nonrelativistic structure and satisfies the empty space Newton gravitational field equations with negative cosmological constant \cite{GiPa}.\footnote{Starting with the group 
\beq
N^-(d)=\big(\SO(d)\times\SO(1,1)^{\uparrow}\big)\ltimes\bbR^{2d}
\label{NH-}
\eeq
we would, similarly, end up with a spacetime with positive cosmological constant.
Here we will focus our attention to $N^+(d)$ in
(\ref{NH+}).}

Below, we extend the above-mentioned construction to our conformal 
Galilei groups $\CGal_N(d)$, at any level $N\geq1$.

\subsection{Conformal Galilei spacetimes}\label{CGalSpaceTimeSection}

The conformal Galilei spacetimes $M_N$, associated with $z=2/N$ where $N=1,2,\ldots$, are introduced by
starting with the conformal Galilei groups $\CGal_{2/z}(d)$, viz.,
\begin{equation}
M_N=\CGal_{N}(d)/H_N
\quad\hbox{where}\quad
H_N=(\rO(d)\times\Aff(1,\bbR))\ltimes\bbR^{Nd}.
\label{MNbis}
\end{equation}
Explicitly, the projection $\pi_N:\CGal_{N}(d)\to{}M_N$ in (\ref{MNbis}) is defined by the direction 
\begin{equation}
\pi_N(\rg)=\bbR^*\cdot\rg_{d+N+1}
\end{equation}
of the last column-vector of the matrix, $\rg$,  in (\ref{CGalN}).
Therefore $x=\pi_N(\rg)$ gets locally identified with
\begin{equation}
\barray{c}
\bx\\ 
t^N\\ 
\vdots\\ 
t\\ 
1
\earray
=
\bbR^*\cdot
\barray{c}
\bB_0\\ 
b^N\\ 
\vdots\\ 
b{}d^{N-1}\\ 
d^N
\earray.
\label{x=pi(g)}
\end{equation}

Intuitively, this amounts to factoring out $\rO(d)$, dilations and expansions and all higher-than-zeroth-order accelerations,
and identifying space-time with ``what is left over''. 

Then, the action $\rg\mapsto\rg_{M_N}$ of~$\CGal_{N}(d)$ on spacetime $M_N$ is  globally given by
$
\rg_{M_N}(\pi_N(h))=\pi_N(\rg h)
$
and, in view of (\ref{x=pi(g)}), retains the 
local form (\ref{x*}) and (\ref{t*}).
Now, by the very definition (\ref{CGalN}) of the conformal Galilei group at level~$N$, we get indeed the conformal Galilei spacetime
\begin{equation}
M_N=(\bbR^d\times\Ver_N(\bbR^2\!\setminus\!\{0\}))/\bbR^*,
\label{MN}
\end{equation}
fibered above the Veronese curve $\Ver_N(\RP^1)\subset\RP^N$, interpreted as the time axis,~$T$.

Besides, it may be useful to view the projective line as $\bbR{}P^1\cong{}S^1/\bbZ_2$, locally parametrized by an angle $\vartheta$ related to the above-chosen affine parameter
\begin{equation}
t=\tan\vartheta,
\label{t=tantheta}
\end{equation}
highlighting that  $\vartheta\sim\vartheta+\pi$. This entails (see (\ref{MN})) that 
\begin{equation}
M_N\cong(\bbR^d\times\Ver_N(S^1))/\bbZ_2
\label{MNtris}
\end{equation}
where $\Ver_N(S^1)$ stands for the image of $S^1$ by the mapping (\ref{uk}).
To summarize, we have the following diagram
\begin{equation}
\begin{CD}
\CGal_{N}(d) 
\strut\\
@V{H_N}VV 
\strut\\
M_N\cong(\bbR^d\times\Ver_N(S^1))/\bbZ_2  @> \bbR^d >> T\cong\Ver_N(\bbR{}P^1)
\end{CD}
\label{diagram}
\end{equation}
where the horizontal arrow denotes the canonical fibration of spacetime $M_N$ onto the time axis $T$.

\goodbreak

Note that for $N=1$ the Schr\"odinger-homogeneous spacetime $M_{1}$, i.e., the  
M\"obius spacetime (\ref{M1}), is obtained.

We will from now on focus our attention to the new $\CGal_{N}(d)$-homogeneous spacetimes $M_N$. 

Let us lastly provide, for the record, explicit formul\ae\ for the projective action $(\bx,\vartheta)\mapsto(\bx^*,\vartheta^*)$ of $\CGal_N(d)$ on our Newton-Hooke manifold~$M_N$, in terms of the angular coordinate introduced in (\ref{t=tantheta}). We will use the local polar decomposition of any element in $\SL(2,\bbR)$, viz.,
\begin{equation}
C=
\left(\begin{array}{rr}
\cos\alpha&\sin\alpha\\[4pt]
-\sin\alpha&\cos\alpha
\end{array}
\right)
\left(\begin{array}{rr}
a&0\\[4pt]
0&a^{-1}
\end{array}
\right)
\left(\begin{array}{rr}
1&0\\[4pt]
c&1
\end{array}
\right),
\label{C}
\end{equation}
where $\alpha\in\bbR/(2\pi\bbZ)$ is a ``time-translation'', $a\in\bbR^*$ a dilation, and $c\in\bbR$ an expansion. Then, Equations (\ref{x*}) and (\ref{t*}) yield
\begin{equation}
\begin{array}{rrll}
A&:&\bx^*=A\bx, &\vartheta^*=\vartheta,
\\[6pt]
\bB_N&:&\bx^*=\bx+\bB_N\tan^N\!\vartheta, &\vartheta^*=\vartheta,\\[6pt]
\vdots&&\vdots&\vdots\\[6pt]
\bB_1&:&\bx^*=\bx+\bB_1\tan\!\vartheta, &\vartheta^*=\vartheta,\\[6pt]
\bB_0&:&\bx^*=\bx+\bB_0, &\vartheta^*=\vartheta,\\[6pt]
\alpha&:&\bx^*=\displaystyle%
\frac{\bx\cos^N\!\vartheta}{\cos^N\!(\vartheta+\alpha)}, &\vartheta^*=\vartheta+\alpha,\\[6pt]
a&:&\bx^*=a^{N}\bx, &\vartheta^*=\arctan(a^{2}\tan\vartheta),\\[6pt]
c&:&\bx^*=\displaystyle%
\frac{\bx}{(c\,\tan\vartheta+1)^N}, &\vartheta^*=
\displaystyle%
\arctan\left(\frac{\tan\vartheta}{c\,\tan\vartheta+1}\right),
\end{array}
\label{NHaction}
\end{equation}
with the same notation as before.

These formul\ae\ extend those derived before, at the Lie algebraic level, for $N=1$, and $N=2$ \cite{Luk07, Galaj}. 
Those in \cite{Galaj}, for example,  are obtained
from  (\ref{NHaction}) by putting $N=2l$ and
introducing, in view of (\ref{x=pi(g)}) and (\ref{t=tantheta}), the new coordinates
\beq
{\bf X}={\bf x}\cos^N\!\vartheta,
\qquad
t=\tan\vartheta.
\label{Xt}
\end{equation}

\subsection{Galilean conformal Cartan connections}\label{CartanConnections}

Prior to  introducing (flat) Cartan connections associated with our conformal Galilei groups, let us recall some basic facts about Galilei connections \cite{Kun}.

\goodbreak

Given a Galilei structure $(M,\gamma,\theta)$ as introduced in Section \ref{NCSection}, we define the bundle of Galilei frames of $M$ as the bundle $P\to{}M$ of those frames 
$(x,e_1,\ldots,e_{d+1})$ such that
\begin{equation}
\sum_{K=1}^d{e_K\otimes{}e_K}=\gamma
\qquad
\text{and}
\qquad
\theta^{d+1}=\theta,
\end{equation}
where $(\theta^1,\ldots,\theta^{d+1})$ is the coframe at $x\in{}M$.

The bundle of Galilei frames is a principal $H$-subbundle of the frame-bundle of $M$, with $H\subset{}G(=\Gal(d))$ the homogeneous Galilei group consisting of those matrices
\begin{equation}
\rh=
\left(
\begin{array}{cc}
A&\bB_1\\
0&1
\end{array}
\right),
\label{H}
\end{equation}
where $A\in\rO(d)$, and $\bB_1\in\bbR^d$. 

\goodbreak
A local coordinate system $(x^\alpha)$ on $M$ (where $\alpha=1,\ldots,d+1$) induces a local coordinate system $((x^\alpha),(e^\alpha_a))$ on $P$, where $e_a=e^\alpha_a\,\partial/\partial{x^\alpha}$ for all $a=1,\ldots,d+1$; with the definition $(\theta^a_\alpha)=(e_a^\alpha)^{-1}$, the $1$-forms
\begin{equation}
\theta^a=\theta^a_\alpha\,dx^\alpha
\label{soldering1form}
\end{equation}
constitute the component of \textit{soldering} $1$-form of~$P$.

Let us denote by $\fg$ (resp. $\fh$) the Lie algebra of $G$ (resp. $H$) so that $\fg=\fh\ltimes\bbR^{d+1}$. A Galilei connection is a $\fg$-valued $1$-form of $P$ such that
\begin{equation}
\omega=
\left(
\begin{array}{cc}
(\omega^a_b)&(\theta^a)\\
0&0
\end{array}
\right),
\label{GalileiConnection}
\end{equation}
where, with the notation of (\ref{galalRep}),
\begin{equation}
(\omega^a_b)=
\left(
\begin{array}{rrr}
\omega&\bbeta_1\\[6pt]
0&0
\end{array}
\right)
\label{HomGalConnection} 
\end{equation}
is an ordinary $\fh$-valued connection $1$-form on the $H$-bundle $P\to{}M$, and 
\begin{equation}
(\theta^a)
=
\left(\begin{array}{c}
\bbeta_0\\[6pt]
\varepsilon
\end{array}
\right)
\label{GalSoldering}
\end{equation}
is the $\bbR^{d+1}$-valued soldering $1$-form (\ref{soldering1form}) of~$P$.\footnote{The translation components of $\omega$ are precisely chosen as those of the soldering $1$-form because Galilei connection are assumed to be affine connections.}

\goodbreak

Then the structure equations provide us with the definition of the associated curvature $2$-form $((\Omega^a_b),(\Omega^a))$ on $P$, namely
\begin{eqnarray}
\label{Omega}
\Omega^a_b&=&\d\omega^a_b+\omega^a_c\wedge\omega^c_b,\\
\label{Theta}
\Omega^a&=&\d\theta^a+\omega^a_c\wedge\theta^c,
\end{eqnarray}
for all $a,b=1,\ldots,d+1$.

Demanding now that the curvature $2$-form be $\fh$-valued (the torsion $(\Omega^a)$ is set to zero), we end up with a symmetric connection $(\Gamma_{\alpha\beta}^\rho)$, entering the following local expression
\begin{equation}
\omega^a_b=\theta^a_\rho(de^\rho_b+\Gamma_{\alpha\beta}^\rho\,dx^\alpha{}e^\beta_b),
\label{GalConnection} 
\end{equation}
such that, if $\nabla$ stands for the associated covariant derivative of spacetime tensor fields, $\nabla_\rho\gamma^{\alpha\beta}=0$ and $\nabla_\alpha\theta_\beta=0$, for all $\alpha,\beta,\rho=1,\ldots,d+1$. This finally entails that $\Gamma$ (given  by $\omega$) is a \textit{Galilei connection} on $(M,\gamma,\theta)$ in the sense of Section~\ref{NCSection}.

At this stage, it is worthwhile mentioning that Galilei connections (\ref{GalileiConnection}) are special instances of ``Cartan connections'' on which the next developments will rely.

Let us thus recall, for completeness, the definition of a \textit{Cartan connection} on a principal fiber bundle $P\to{}M$, with structural group a closed subgroup $H$ of a Lie group $G$, where $\dim(M)=\dim(G/H)$. Put $\fg=\mathrm{Lie}(G)$ and $\fh=\mathrm{Lie}(H)$. 

\goodbreak

Such a ``connection'' is given by a $\fg$-valued $1$-form $\omega$ on the principal $H$-bundle~$P$ such that 
\begin{enumerate}
\item
$\omega(Z_P)=Z$ for all $Z\in\fh$
\item
$(h_P)^*\omega=\Ad(h^{-1})\omega$ for all $h\in{}H$
\item
$\ker\omega=\{0\}$
\end{enumerate}
where the subscript $P$ refers to the group or Lie algebra right-action on~$P$. 
These connections provide a powerful means to encode the geometry of manifolds modeled on homo\-geneous spaces $G/H$, e.g., projective or conformal geometry. 

We will show, in this section, that the homogeneous spaces $M_N$ (see (\ref{MN})) indeed admit, for all $N=1,2,\ldots$, a conformal Newton-Cartan structure together with a distinguished, \textit{flat}, normal Cartan connection 
associated with $\CGal_N(d)$.\footnote{Since we are dealing here with homogeneous spaces $G/H$, it will naturally be given by the (left-invariant) Maurer-Cartan $1$-form of the corresponding groups $G$.}
The general construction of the normal Cartan connection associated with a Schr\"odinger-conformal Newton-Cartan structure has been performed in \cite{Duv0}.

\goodbreak

$\bullet$ The Galilei group $\Gal(d)$ can be viewed as the bundle of Galilei frames over spacetime
$\bbR^d\times\bbR=\Gal(d)/(\rO(d)\ltimes\bbR^d)$,
cf. (\ref{GalST}). 
Using (\ref{Gal}), we find that the (left-invariant) Maurer-Cartan $1$-form $\Theta_{\Gal(d)}=\rg^{-1}\d\rg$ reads
\begin{equation}
\Theta_{\Gal(d)}=
\left(
\begin{array}{rrr}
\omega&\bbeta_1&\bbeta_0\\[6pt]
0&0&\varepsilon\\[6pt]
0&0&0
\end{array}
\right),
\label{GalMC} 
\end{equation}
where $\bbeta_0$, and $\varepsilon$ are interpreted as the components of the soldering {$1$-form (\ref{soldering1form}) of the principal $H$-bundle $\Gal(d)\to\bbR^d\times\bbR$. Then, the Maurer-Cartan structure equations $\d\Theta+\Theta\wedge\Theta=0$ read
\begin{eqnarray}
\label{domega}
0&=&\d\omega+\omega\wedge\omega,\\
\label{dbeta1}
0&=&\d\bbeta_1+\omega\wedge\bbeta_1,\\
\label{dbeta0}
0&=&\d\bbeta_0+\omega\wedge\bbeta_0+\bbeta_1\wedge\varepsilon,\\
\label{depsilon}
0&=&\d\varepsilon.
\end{eqnarray}
Clearly, the Maurer-Cartan $1$-form (\ref{GalMC}) endows the bundle $\Gal(d)\to\bbR^d\times\bbR$ with a Cartan connection in view of the above defining properties of the latter. This connection is canonical and \textit{flat}.
Indeed, Equations (\ref{domega}, \ref{dbeta1}), specializing~(\ref{Omega}), entail that the connection $2$-form $(\omega,\bbeta_1)$ is flat, while Equations (\ref{dbeta0}, \ref{depsilon}), 
cor\-responding to $(\Omega^a)=0$ in (\ref{Theta}), 
guarantee having zero torsion.
 See, e.g., \cite{Kun,Duv0}. 
We note that, 
with the standard notation used throughout our article,
the Galilei structure is given here by $\gamma=\delta^{ij}A_i\otimes{}A_j$, where $A\in\rO(d)$ represents an orthonormal frame, together with the clock $1$-form $\theta=\varepsilon$; see (\ref{Gal}).
At last, in view of the general form~(\ref{GalConnection}) of Galilei connections, there holds
\begin{equation}
\Gamma^\rho_{\alpha\beta}=0
\label{flatGamma} 
\end{equation}
for all $\alpha,\beta,\rho=1,\ldots,d+1$, in the spacetime coordinate system $(x^i=B_0^i, x^{d+1}=b)$ provided by the matrix realization (\ref{Gal}) of $\Gal(d)$.

\goodbreak

$\bullet$ Likewise, if $N=1$, the Schr\"odinger group $\Sch(d)$ may be thought of as a subbundle of the bundle of \emph{$2$-frames} of spacetime $M_1\cong\Sch(d)/H_1$, see (\ref{MNbis}).\footnote{The bundle of $2$-frames is called upon since the vector fields (\ref{schd}) spanning $\sch(d)$ are polynomials of degree $2$ in the spacetime coordinates.} This time, $\Sch(d)$ is, indeed, interpreted as the bundle of conformal Galilei $2$-frames associated with the conformal class $\gamma\otimes\theta$ of a Galilei structure $(\gamma,\theta)$ over $M_{1}$ 
as given by Equation (\ref{galconfMN}) with $m=n=1$. The Maurer-Cartan $1$-form $\Theta_{\Sch(d)}$ of this group actually gives rise to the canonical flat Cartan connection on the $H_1$-bundle 
$ 
\Sch(d)\to{}M_{1}.
$
We refer to \cite{Duv0,BDP} for a comprehensive description of Schr\"odinger conformal Cartan connections. 

\goodbreak

Using the same notation as in (\ref{sch2dRep}), we find
\begin{equation}
\Theta_{\Sch(d)}=
\left(
\begin{array}{rrr}
\omega&\bbeta_1&\bbeta_0\\[6pt]
0&\lambda&\varepsilon\\[6pt]
0&-\kappa&-\lambda
\end{array}
\right)
\label{SchMC} 
\end{equation}
and, hence, the structure equations
\begin{eqnarray}
\label{domegaSch}
0&=&\d\omega+\omega\wedge\omega,\\
\label{dbeta1Sch}
0&=&\d\bbeta_1+\omega\wedge\bbeta_1-\bbeta_0\wedge\kappa,\\
\label{dbeta0Sch}
0&=&\d\bbeta_0+\omega\wedge\bbeta_0+\bbeta_1\wedge\varepsilon-\bbeta_0\wedge\lambda,\\
\label{delambdaSch}
0&=&\d \lambda-\varepsilon\wedge \kappa,\\
\label{depsilonSch}
0&=&\d\varepsilon-2\varepsilon\wedge\lambda,\\
\label{dekappaSch}
0&=&\d\kappa-2 \kappa\wedge\lambda.
\end{eqnarray}

Furthermore, a Galilei structure can be introduced on spacetime $M_1$ viewed as the quotient of  (\ref{NH+}).
Our clue is that embedding $\SO(2)$ into $\SL(2,\bbR)$ through
\begin{equation}
\vartheta\mapsto
\left(\begin{array}{rr}
\cos\vartheta&\sin\vartheta\\
-\sin\vartheta&\cos\vartheta
\end{array}
\right),
\label{SO2}
\end{equation}
the group
\begin{equation}
\New_1(d)=(\rO(d)\times\SO(2))\ltimes\bbR^{2d}
\label{New1}
\end{equation}
we will call the \emph{Newton-Hooke group of level $N=1$} in what follows,\footnote{Its neutral component is $N^{+}(d)$ in (\ref{NH+}).}
becomes a \emph{subgroup} of 
the Schr\"odinger group,
\beq
\New_1(d)\subset \Sch(d).
\eeq
Then, $\New_1(d)$ can readily be identified with the bundle of Galilei frames of 
\beq
M_1\cong\New_1(d)/(\rO(d)\times\bbZ_2)\ltimes\bbR^d.
\label{NH1manif}
\eeq

Introducing the pulled-back Maurer-Cartan $1$-form, $\Theta_{\New_1(d)}$, we end up with the previous structure equations (\ref{domegaSch})--(\ref{dekappaSch}) specialized to the case $\lambda=0$, and $\kappa=\varepsilon$. Comparison between the latter equations, and the (flat) Galilei structure equations shows that both sets coincide except for the equations characterizing $\d\bbeta_1$. This entails that $M_{1}$ acquires curvature through the $2$-form \cite{Duv0}
\begin{equation}
\bOmega_1=\bbeta_0\wedge\varepsilon.
\label{Omega1}
\end{equation}

Let us present, for completeness, a local expression of the NC-Cartan connection on~$M_1$ generated by $\Theta_{\New_1(d)}$. Putting $\bx=\bB_0$ defines, along with $\vartheta\in\bbR/(2\pi\bbZ)$ introduced in (\ref{SO2}), a co\-ordinate system on spacetime $M_1$. 
We readily find the components of the soldering $1$-form to be 
\begin{equation}
\bbeta_0=A^{-1}(\d\bx-\bB_1\d\vartheta)
\qquad
\text{and}
\qquad
\varepsilon=\d\vartheta,
\end{equation}
see (\ref{GalSoldering}), while those of the $\fh$-connection write 
\begin{equation}
\omega=A^{-1}\d{}A
\qquad
\text{and}
\qquad
\bbeta_1=A^{-1}(\d\bB_1+\bx\,\d\vartheta),
\end{equation}
see~(\ref{HomGalConnection}).
From the last expression we infer 
(exploiting the general form (\ref{GalConnection}) of Galilei connections) 
that the only nonzero components of the Christoffel symbols of the connection are
\begin{equation}
\Gamma^{i}_{\vartheta\vartheta}=x^i
\label{Gamma}
\end{equation}
for all $i=1,\ldots,d$. This entails that the nonzero components of curvature tensor $R_1$, associated with~$\bOmega_1$, are
$
(R_1)^i_{j\vartheta\vartheta}=\delta^i_j
$
for all $i,j=1,\ldots,d$. 
This Galilei connection is clearly a NC-connection since (\ref{SymCurv}) holds true.

 The only nonvanishing component of the Ricci tensor is therefore 
\beq
(\Ric_1)_{\vartheta\vartheta}=d.
\label{Riccitensor}
\eeq
 Hence, our NC-connection $(\omega,\bbeta_1)$ provides us with a solution of the NC-field equations (\ref{NCfieldEqs}), 
\begin{equation}
\Ric_1+\tLambda_1\theta\otimes\theta=0,
\label{Ric1}
\end{equation}
where $\theta=\d\vartheta$; the ``cosmological constant'' 
\begin{equation}
\tLambda_1=-d
\label{Lambda1}
\end{equation}
is therefore given by the \emph{dimension of space}.

The angular parameter $\vartheta$ being dimensionless, it might be worth intro\-ducing, at this stage, a (circular) \emph{time parameter} 
\begin{equation}
\tau=H_0^{-1}\vartheta,
\label{tau}
\end{equation}
where $H_0$ is a ``Hubble constant'' whose inverse would serve as a time unit.\footnote{Unlike in  general relativity, no specific parameter such as the de Sitter spacetime radius is available in our nonrelativistic framework; whence this somewhat arbitrary choice of a time unit.} This entails that $(\Ric_1)_{\tau\tau}=H_0^2d$ (compare (\ref{Riccitensor})), and, hence, provides us with the cosmological constant
\begin{equation}
\Lambda_1=-H_0^2d.
\label{Lambda1bis}
\end{equation}
Now, cosmologists usually introduce the \emph{reduced cosmological constant}, $\lambda$, in terms of the cosmological constant, $\Lambda$, the Hubble constant, $H_0$, and the dimension of space, $d$, via 
\begin{equation}
\Lambda=\lambda{}H_0^2d.
\label{deflambda}
\end{equation}
In view of (\ref{Lambda1bis}), our model therefore yields a reduced cosmological constant 
\begin{equation}
\lambda_1=-1.
\label{lambda1}
\end{equation}

\goodbreak
$\bullet$ For the case $N>1$, the Maurer-Cartan $1$-form of $\CGal_N(d)$ reads
\begin{equation}
\Theta_{\CGal_N(d)}=
\left(
\begin{array}{rrrrrr}
\omega&\bbeta_N&\bbeta_{N-1}&\cdots&\bbeta_1&\bbeta_0\\[6pt]
0&N\lambda&N\varepsilon&\cdots&0&0\\[6pt]
0&-\kappa&(N-1)\lambda&\cdots&0&0\\[6pt]
\vdots&\vdots&\ddots&\ddots&0&0\\[6pt]
0&0&\cdots&\cdots&2\varepsilon&0\\[6pt]
0&0&\cdots&\cdots&(1-N)\lambda&\varepsilon\\[6pt]
0&0&0&0\,\cdots\,0&-N\kappa&-N\lambda
\end{array}
\right)
\label{CGalMC} 
\end{equation}
with the same notation as before. 

Then, the preceding computations can be reproduced, \textit{mutatis mutandis}, for the group~$\CGal_{N}(d)$ which serves as the bundle of conformal Galilei $N$-frames of~$M_N$. 
At that point, as a straightforward generalization of (\ref{New1}), we can define the \emph{Newton-Hooke group at level $N$} as 
\beq
\New_N(d)=(\rO(d)\times\SO(2))\ltimes\bbR^{d(N+1)}\subset\CGal_N(d).
\label{NHN}
\eeq

Much in the same manner as in the case $N=1$, the NC-connection obtained from the pull-back $\Theta_{\New_N(d)}$ of the Maurer-Cartan $1$-form (\ref{CGalMC}) can be computed. It is easily shown that the curvature of the homogeneous spacetimes $M_N$ is now given by
\begin{equation}
\bOmega_N=N\bOmega_1.
\label{OmegaN}
\end{equation}
This connection turns out to produce an exact solution of the vacuum  NC-field equations~(\ref{NCfieldEqs}), the reduced cosmological constant at level $N$ being now given by
\begin{equation}
\lambda_N=-N
\label{lambdaN}
\end{equation}
for all $N=1,2,\ldots$.

\goodbreak

\section{Conclusion and outlook}

Our main results are two-fold: firstly, we have
found that the infinite-dimensional Lie algebra of infinitesimal conformal Galilei transformations with rational dynamical exponent admits, in fact, finite dimensional Lie subalgebras provided the
dynamical exponent is $z=2/N$, where $N=1, 2,\dots$.
Then, we have proposed a natural construction devised to integrate, for each $N$, these Lie algebras into Lie groups, named \emph{Conformal Galilei groups at level $N$}, by means of the  classic Veronese embeddings~\cite{OT}. 
The values $N=1$ and $N=2$ correspond to the Schr\"odinger Lie algebra, and to the  CGA, respectively.
Our results are somewhat unexpected in that, starting with the symmetry problem in
the Galilean context, we end up with Newton-Hooke space-times and their symmetries.

Let us shortly list some of the  applications beyond the by-now standard $N=1$, i.e., $z=2$ Schr\"odinger symmetry.

The $N=2$, i.e., $z=1$ conformal extension of the [exotic] Galilean algebra was
studied in \cite{SZ03}, and later extended so as to include also accelerations \cite{LSZGalconf}.  

Newton-Hooke symmetry and spacetime have been considered in nonrelativistic cosmology \cite{BaLL,Aldo,GiPa,Pekin,Luk07,Arratia}. It has been
proved \cite{Saka}  that the $N=1$
 Galilean Conformal algebra is isomorphic
to the Newton-Hooke string algebra
studied in string theory~\cite{Gomis}. 

\goodbreak

The values $N=4$, and $N=6$, i.e., the dynamical exponents $z=1/2$ and $z=1/3$ arise
in statistical mechanics, namely for the spin-spin correlation function in the axial 
next-nearest-neighbor spherical model at its Lifschitz points of first and second order~\cite{Hen97,Hen02}. 

Our Conformal Galilei groups, $\CGal_N(d)$, which generalize the Schr\"odinger ($N=1$) and the Conformal Galilei ($N=2$) cases 
to any integer $N$, do \emph{not} act regularly on ordinary flat Galilean spacetime, however; they act rather on manifolds constructed from them,
which are analogous to the conformal compactification of Minkowski spacetime. Moreover,
their group structure allows us to recover
these spacetimes as homogeneous spaces 
 of the Newton-Hooke groups $\New_N(d)$ as defined in (\ref{NHN}), and generalizing 
to level $N$ the Newton-Hooke group (\ref{NH+}).
These as\-sociated spacetimes $M_N$ are endowed with a (conformal) Newton-Cartan structure by construc\-tion. Remarkably,
they are identified as \emph{Newton-Hooke spacetimes
with quantized negative reduced cosmological constant},
$\lambda_N$ in~(\ref{lambdaN}).\footnote{Spacetimes with positive cosmological constant
would be obtained by replacing $\SO(2)$ by $\SO(1,1)^{\uparrow}$ as in (\ref{NH-}); see also \cite{GiPa}.}

An intuitive way of understanding our strategy is to consider, say, Schr\"odinger expansions  in (\ref{SchExpan}), or in (\ref{SchAction}), and
observe that it is the \emph{denominator} 
which makes the group action singular.
Our way of removing this ``hole'' singularity} is factorize
the denominator in these expressions, as dictated by the projective action in~(\ref{projAction}).

\goodbreak

Having constructed our finite-dimensional
conformal extensions for each $N$, there remains the task to find 
\emph{physical realizations}.

Firstly, in the Schr\"odinger case $N=1$, one
can verify directly that the geodesic equations, i.e., the
\emph{free Newton equations},
\beq
\frac{d^2\vx}{dt^2}=0,
\label{freeNewtoneq}
\eeq
are, indeed, preserved due to the following transformation of the acceleration
\begin{equation}
\frac{d^2{\vx}^*}{dt^{*2}}=A\frac{d^2\vx}{dt^2}\frac{dt}{dt^*},
\label{accelerationTransfo} 
\end{equation}
where $A\in\rO(d)$.
Similarly, the equations of \emph{uniform acceleration},
\beq
\frac{d^3\vx}{dt^3}=0,
\label{unifaccel}
\eeq
are preserved by the conformal Galilei group, $\CGal_1(d)$, in view of the following transformation law, namely,
\begin{equation}
\frac{d^3{\vx}^*}{dt^{*3}}=A\frac{d^3\vx}{dt^3}\left(\frac{dt}{dt^*}\right)^2.
\label{suraccelerationTransfo} 
\end{equation}
These examples make it plausible 
that the $N$-conformal symmetry, $\CGal_N(d)$, is realized
by the
higher-order geodesic equations,
\beq
\frac{d^{N+1}\vx}{dt^{N+1}}=0.
\label{unifaccelN}
\eeq
Such a statement is suggested by the quantum formulas in \cite{Hen97,Hen02} and is consistent, for $N=2$,
with Ref. \cite{Fedoruk}.
Another promising approach is  \cite{GalaP}.

We would finally like to mention that a complete, general, construction of Cartan connections for conformal Galilei structures, modeled on the $\CGal_N(d)$-homogeneous manifolds~$M_N$, still remains to be undertaken in order to extend that of normal Schr\"odinger-Cartan connections carried out in \cite{BDP}. 
This would lead to a satisfactory, brand-new, geometric definition of the group of $N$-conformal Galilei transformations of a NC-structure as the group of automorphisms of such an as\-sociated normal Cartan connection.

Let us now discuss the relationship of our procedures and technique to some other work on the same subject, namely to conformal Galilean symmetries.

Gibbons and Patricot
\cite{GiPa} derive the Newton-Cartan structure
of Newton-Hooke spacetime in the ``Bargmann'' framework of Ref. \cite{DBKP-DGH}. This
null ``Kaluza-Klein-type'' approach provides us indeed with a
preferred way of defining a nonrelativistic
structure  ``downstairs''.
A similar explanation holds in a uniform
magnetic field
\cite{BDPLMP,DHP94,HaHo00,GiPo,Alvarez,ZHKohn}.

\goodbreak

No central terms are considered in
this paper: 
our Lie algebras are represented by vector fields on 
(flat) Newton-Cartan spacetime.
Central terms, and the mass in particular, are 
important, though, and are indeed necessary for physical applications.
Henkel \cite{Hen97, Hen02} does actually consider mass terms: he works with \emph{operators} with such terms involving $c^{-2}$,
where $c$ is the speed of light. In his approach, those
appear in the coefficients of \emph{powers} of $\partial/\partial{t}$.  Considering higher-order 
terms in powers of $\partial/\partial{t}$ goes beyond our framework, though.
For  finite~$c$, Henkel's boosts are Lorentzian, 
not Galilean, however; our center-free Lie algebras with genuine Galilei boosts are
recovered as $c\to\infty$, yielding  also $(N-1)$ accelerations in addition. At last, the mass terms disappear, as they should: mass and accelerations are indeed inconsistent \cite{LSZGalconf}.

\goodbreak

Central extensions of the conformal Galilei algebra have been considered in \cite{LSZGalconf,Alvarez} in the planar case, $d=2$, and in \cite{NOR-M} for $d=3$. 

Let us finally mention a natural further program, in the wake of the present study, namely the determination of the group-co\-homologies of our Conformal Galilei groups. Also would it be worthwhile to classify all symplectic homogeneous spaces \cite{Sou} of the latter, likely to unveil new, physically interesting,  systems.

\kikezd{Acknowledgments}:
P.A.H is indebted to the \textit{Institute of Modern Physics} of the Lanzhou branch of the Chinese Academy of Sciences for hospitality extended to him, 
and to P.-M. Zhang for discussions. Useful correspondence is acknowledged to A. Galajinsky, J. Gomis, M. Henkel, J. Lukierski, M. Plyushchay, as well as enlightening discussions to S. Lazzarini. 


\end{document}